\documentstyle[12pt,amstex,psfig,amssymb,righttag]{article}
\begin{document}
\renewcommand{\thefootnote}{\fnsymbol{footnote}}

\thispagestyle{empty}

\vspace*{-1cm}
\begin{center}
{\Large \bf Some Comments on Wheeler De Witt Equation for
Gravitational Collapse and the Problem of Time}

\vspace{3mm}
by\\
\vspace{3mm}
{\sl Carlos
Pinheiro$^{\ddag}$\footnote{fcpnunes@@cce.ufes.br/maria@@gbl.com.br}
} 
and 
{\sl F.C. Khanna$^{+}$\footnote{khanna@@phys.ualberta.ca}}

\vspace{3mm}
$^{\ddag}$Universidade Federal do Esp\'{\i}rito Santo, UFES.\\
Centro de Ci\^encias Exatas\\
Av. Fernando Ferrari s/n$^{\underline{0}}$\\
Campus da Goiabeiras 29060-900 Vit\'oria ES -- Brazil.\\

$^{+}$Theoretical Physics Institute, Dept. of Physics\\
University of Alberta,\\
Edmonton, AB T6G2J1, Canada\\
and\\
TRIUMF, 4004, Wesbrook Mall,\\
V6T2A3, Vancouver, BC, Canada.
\end{center}

\vspace{3mm}
\begin{center}
Abstract
\end{center}

We write the Hamiltonain for a gravitational spherically symmetric
scalar field collapse with massive scalar field source, and we
discuss the application of Wheeler De Witt equation as well as the
appearence of time in this context. Using an Ansatz for Wheeler
De Witt equation, solutions are discussed including the appearence
of time evolution.

\newpage
\section*{Introduction}
\setcounter{footnote}{0}
\paragraph*{}

In this letter we discuss the problem of gravitational collapse of
a star using the Wheeler-De Witt equation.

In  accordance with \cite{dois} we assume a scalar field, $\phi$,
with a mass term and we assume that the super hamiltonian has a
constraint [1-5] such that $H\simeq 0$. Ordering of
operators is assumed.  

A particular ansatz for the functional is chosen to show
qualitatively the appearance of the notion or concept of ``time''
after quantization.

As in the case of the hydrogen atom the discrete index is identified
with an ``internal time'' just as in any relativistic field theory or
general relativity but different from the usual quantum mechanics,
where ``time'' appears as a Galilean time.

We apply the Wheeler-De Witt equation for a special collapse
condition despite the fact that the question related to the
Copenhagen interpretation for product of functional $\psi (\Lambda
,R,\phi )$ is not understood.

Let us begin by writing the super Hamiltonian for a gravitational
spherically symmetric scalar field collapse with massive scalar field
source such as \cite{dois}.
\begin{equation}
H={\cal H}+\frac{1}{2}\ m^2R^2\Lambda \phi^2\ , 
\end{equation}
where
\begin{eqnarray}
{\cal H} &=& -R^{-1}P_RP_{\Lambda}+\frac{1}{2}\ R^{-2}\Lambda P^2_{\Lambda}+
\Lambda^{-1}RR''-\Lambda^{-2}RR'\Lambda '+\frac{1}{2}\ \Lambda^{-1}
R{'}^2+\nonumber \\
&-&\frac{1}{2}\ \Lambda +\frac{1}{2}\ R^{-2}\Lambda^{-1}P^2_{\phi}+
\frac{1}{2}\ R^2\Lambda^{-1}\phi{'}^2\ .
\end{eqnarray}

In the expression above $P_R,P_{\Lambda},P_{\phi}$ imply respectively
conjugate momenta associated with $R,\Lambda$ and $\phi$ variables.

Furthermore $R=R(r,t)$, $\Lambda =\Lambda (r,t)$, $\phi =\phi (r,t)$.
We define conjugate momentum as 
\begin{equation}
\pi_x=-i\ \frac{\partial}{\partial x}
\end{equation}
where $\underline{x}$ means $R,\Lambda$ or $\phi$ variable.

It is a known fact that using the Hamiltonain (2) some 
operator ordering problems appear \cite{um,dois}.

A simple form to represent the ambiguous order of factors 
$\left(x\ , \ \displaystyle{\frac{\partial}{\partial x}}\right)$ and 
$\left(y\ , \ \displaystyle{\frac{\partial}{\partial y}}\right)$ is
given by \cite{um}. Applying such an  ordering for operators in (2)
we can find the following squared conjugate momenta 
\begin{eqnarray}
\pi^2_x &=&-\frac{\partial^2}{\partial x^2}-\frac{p}{x}\ 
\frac{\partial}{\partial x}\nonumber \\
\\
\pi^2_y &=&-\frac{\partial^2}{\partial y^2}-\frac{q}{y}\ 
\frac{\partial}{\partial y}\nonumber 
\end{eqnarray}
where $(p,q)$ are $c$-numbers.

It is assumed that the Hamiltonian (2) is a
constraint for a classical Hamiltonain with the mass term present
for the scalar field $\phi$. In other words, the canonical quantization
needs the annihilation of the wave function $\psi$ by the
corresponding quantum operator 
\begin{equation}
\hat{H}\psi =0
\end{equation}
that results in the Wheeler-De Witt equation. Using eq. (2-5) we get
\begin{equation}
\frac{\Lambda}{2R^2}\left(\frac{\partial^2\psi}{\partial \Lambda^2}+
\frac{p}{\Lambda}\ \frac{\partial \psi}{\partial \Lambda}\right)+
\frac{1}{2R^2\Lambda}\left(\frac{\partial^2\Lambda}{\partial \phi^2}+
\frac{q}{\phi}\frac{\partial \psi}{\partial \phi}\right)-
\frac{1}{R}\ \frac{\partial^2\psi}{\partial R\partial \Lambda}\equiv
V \psi
\end{equation}
where $\psi$ is a functional of $\Lambda$, $\phi$ and $R$ functions,
and $V$ is a potential term written as
\begin{equation}
V=\frac{R}{\Lambda}\ R''-\frac{R}{\Lambda^2}\ R'\Lambda '+\frac{1}{2\Lambda}\
R{'}^2-\frac{1}{2}\ \Lambda +\frac{1}{2}\ \frac{R^2}{\Lambda}\ \phi{'}^2 +
\frac{1}{2}\ m^2R^2\Lambda \phi^2
\end{equation}

The prime means derivative with respect to the coordinate
$\underline{r}$. Observe that in equation (6) we don't have any
derivative with respect to time. This means that the equation (6)
could be describing a spherically symmetric gravitational collapse but
without any explicit time dependence for functional $\psi$. The
concept of ``time'' in this case may appear only after quantization in
accordance with \cite{tres}.

This suggests that eq. (6) is like the usual Schr\"{o}dinger equation of
quantum mechanics applied to gravitational collapse but with a
difference depending on the operator ordering [1-5].

The usual Schr\"{o}dinger equation is written as 
\begin{equation}
H\psi =i\frac{\partial \psi}{\partial t}
\end{equation}
where $H$ means the Hamiltonian of the system. It means that
the wave function of the system has an important difference with
equation (6) besides the fact that $\psi$ in (8) to be a function
while $\psi$ in (6) being a functional $\psi (\Lambda ,\phi ,R)$.
The parameter ``time'' $\underline{t}$ in (8) is a universal
time-``external time'' in the sence of Galili-Newton time, while in
equation (6) ``time'' is an internal parameter. In some sense there
is no ``time'' with which we could describe the evolution of
gravitational collapse of the star for exemplo. Thus, in principle we
might apply the equation for a static case such as Schwarszchild
solution but not for a dynamic case where the functions $R,\Lambda
,\phi$ might be time dependent. In other words, one can apply
Wheeler-De Witt equation (6) for static Schwarszchild case where
$R=R(r)$, $\Lambda =\Lambda (r)$ and $\phi =\phi (r)$ but shall we
apply the same equation for the general case, with $R=R(r,t)$, 
$\Lambda =\Lambda (r,t)$ and $\phi =\phi (r,t)$?

How does the conception of ``time'' appear in this case?

How can we get the notion of evolution in time of a collapsing star
using equation (6) without explicit time dependence of the functional $\psi$?

The equation (8) can be applied for steady systems such as hydrogen
atom where the right side is zero and we have
\begin{equation}
\hat{H}\psi =E\psi =0
\end{equation}
where $E$ is the energy. In the particular case of $E=0$ this equation
has a strong resemblance to the Wheeler-De Witt equation.

It is a well known fact that stationary solution can be find from
equation (9) in terms of $R(r),\ \Theta (\theta ),\ \phi (\varphi )$
with $R$, the radial solution and $\Theta (\theta )\phi (\varphi
)=Y(\theta ,\varphi )$  being the spherical harmonics. The obvious
similarity of eq. (9) and eq. (5) leads us to think that eq. (6) can
be solved in the general case, with an ``internal time'' and the idea
of ``evolution'' being identified with some discrete index
$i=1,2,3\cdots $. after solving eq. (6). 

We know that there are many different $\psi_{k\ell m}(r,\theta
,\varphi )$ for different values of $k,\ell ,m$ for the hydrogen
atom and in some sense ``the evolution of the system'' can be seen
as a changing of wave function for a stationary situation. There is
no ``external time'' in eq. (8) for the hydrogen atom.

In the same way we can think of applying in eq. (6) with an ``internal
time'' or without an external time any way and to obtain the
functional $\psi (\Lambda ,\phi ,R)$.

We may take an appropriate ansatz for the eq. (6) and to verify
if it really does satisfy eq. (6). But immediately two questions can be
raised. 

First, which ansatz? There are an inifinite number of possibilities.

Second, the introduction of  a mass term in (1) for scalar field
$\phi$ can break the ``constraint'' character for $H$ and eq. (5)
may not be valid anymore. We must remind that we are assuming the
presence of mass of the scalar field and it does not break the constraint
of super Hamiltonian as in \cite{dois}.

In general the Wheeler-De Witt equation can be separated depending on
the potential term (7). The role of 
$V\left(R,R',R'',\Lambda ,\Lambda ',\phi ,\phi ',m\right)$ is similar
to the coordinates system for decoupling of the Schr\"{o}dinger
equation. It is a known fact that the Schr\"{o}dinger Equation can be separated
in several coordinates systems. In the same manner eq. (6) may decouple
for $\psi (R,\Lambda ,\phi )$ depending on the potential term and the
particular choice of the {\sl ansatz} for the $\psi$
functional. But eq. (6-9) is too complicated and again there is no
derivative in ``time''.

Qualitatively the problem can be solved in the following way. Suppose
that $\psi$ functional reads as 
\begin{equation}
\psi \left(\Lambda ,\phi ,R\right)=\Lambda (r+c)
\sqrt{\phi (r+c)}\ R(r+c)
\end{equation}
where $\Lambda ,R,\phi$ are functions of  $\ \underline{r}\ $ only since there
is no``external time'' as in eq. (8) or an ``internal time'' as in
general Relativity theory or in the relativistic Klein Gordon equation.

In eq. (9) $\underline{c}$ is a constant that can be identified with
``time'' after quantization.

A class of solutions such as is shown below may be found 

\vspace{3mm}
\centerline{\psfig{figure=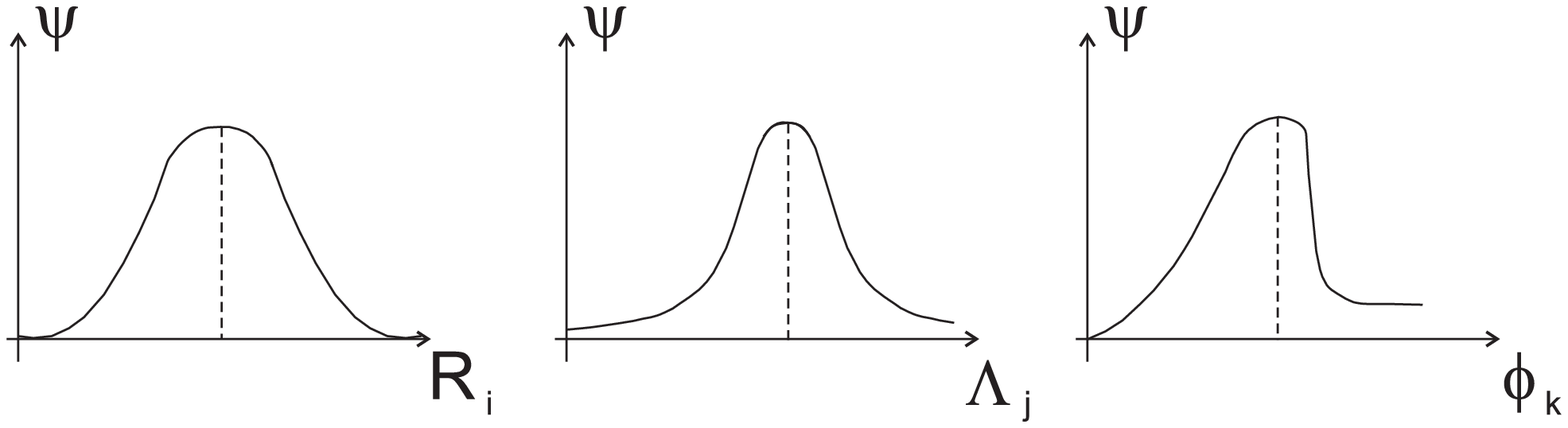,height=3cm}}
\vspace{3mm}

In reality we can find a sequence of $R_i(r),\ \Lambda_j(r)$ and
$\phi_k(r)$ where $i,j,k=1,2,3\cdots$ the concept of ``time'' being
identified with $i,j,k\sim t$ (Time).

In the Schr\"{o}dinger equation for the hydrogen atom the wave
function $\psi_{k\ell m}(r,t)$ can be written as a product of 
$R_{k\ell}(r),\Theta_{\ell m}(\theta )$ and $\phi_m(\varphi )$ for stationary
states and one may see a notion of ``evolution'' through the different
configurations is possible given by different values of $k,\ell ,m$. 

In our case the same idea can be utilised by identifying with a
discrete 
index $(i=1,2,3\cdots )$ as the ``time'' where $i,j,k$ are the
different functions that contribute to our functional $\psi$.

Finally, we need to be clear that eq. (6) has a infinite number of solutions
with the proposal given by eq. (10) being one of them.

The Wheeler-De Witt equation itself has many different
possibilities depending of the operator ordering
\cite{um,dois,tres,quatro,cinco}. Then, in principle one can write
different mathematics (different Wheeler-De Witt equations) and each
one of them with infinite number of ansatz. Each possibility is given
us a notion of ``Time'' after quantization.

The natural question that we can put is:

Shall we find the same ``physics'' for different Wheeler-De Witt equations?

Can we find the same notion of ``time'' from different Wheeler-De
Witt equation with infinite possibilities of the ansatz ?

The physical ``time'' is the same for each possibility or do we have
many times in physics as in \cite{seis}?

Admitting that our equation (6) has some meaning and that the ansatz
eq. (10) can provide us with a notion of ``time'' arises from the
discretization of the index $i,j,k\sim t$. The next question we need to
resolve is: if eq. (6) implies the Schr\"{o}dinger equation for
a global Universe in general and in our particular case it is a
Schr\"{o}dinger equation for a gravitational collpse of a body like
a star how can we improve the Copenhagen interpretation for the
functional $\psi (\Lambda ,\phi ,R)$?

Maybe the answer can be found  as in eq. (6) and the ansatz given by
eq. (10)
describing the possibility of finding the star between
$\underline{m}$ and $\underline{m+dm}$ mass states.

But if so, can it be supported by the condition $m\neq 0$ for the
scalar field $\phi$ in (1)?

Should the superhamiltonian be a real constraint $H\sim 0$ on that condition?

In any case we need to understand the real meaning of operator
ordering in quantum mechanics as well as the meaning of time in all
of physics. While we don't know the final answer for these open
questions there have been uncertain consequences for a complete
understanding of physics and our interpretation for the world.

\subsection*{Acknowledgements:}

\paragraph*{}
I would like to thank the Department of Physics, University of
Alberta for their hospitality. This work was supported by CNPq
(Governamental Brazilian Agencie for Research.

I would like to thank also Dr. Don N. Page for his kindness and attention
with  me at Univertsity of Alberta.

\end{document}